\documentclass[5p, twocolumn]{elsarticle}

\usepackage{upgreek} 
\usepackage{amsmath} 
\usepackage{amssymb} 
\usepackage{amsfonts} 
\usepackage{multirow} 
\usepackage{color} 
\usepackage{float} 

\date{}

\begin{document}

\begin{frontmatter}
\title{\textcolor{black}{3D microstructural generation from 2D images of cement paste \\using generative adversarial networks}}

\author[mysecondaryaddress]{Xin Zhao}
\author[mymainaddress,myfouraddress]{Lin Wang\corref{mycorrespondingauthor}}
\cortext[mycorrespondingauthor]{Corresponding author: Lin Wang (wangplanet@gmail.com)}
\author[mysecondaryaddress,mythirdaddress]{Qinfei Li}
\author[mysecondaryaddress,mythirdaddress]{Heng Chen}
\author[mymainaddress]{Shuangrong Liu}
\author[mythirdaddress]{Pengkun Hou}
\author[CBMA]{Jiayuan Ye}
\author[SD]{Yan Pei}
\author[mysecondaryaddress]{Xu Wu}
\author[mymainaddress]{Jianfeng Yuan}
\author[mymainaddress,myfouraddress]{Haozhong Gao}
\author[myfouraddress,mymainaddress]{Bo Yang}

\address[mysecondaryaddress]{Shandong Provincial Key Laboratory of Preparation and Measurement of Building Materials, \\University of Jinan, Jinan 250022, China}
\address[mymainaddress]{Shandong Provincial Key Laboratory of Network Based Intelligent Computing, \\University of Jinan, Jinan 250022, China}
\address[myfouraddress]{Quan Cheng Laboratory, Jinan 250100, China}
\address[mythirdaddress]{School of Materials Science and Engineering, University of Jinan, Jinan 250022, China}
\address[CBMA]{{State Key Laboratory of Green Building Materials, China Building Materials Academy, Beijing 100024, China}}
\address[SD]{{School of Civil Engineering, Institute of Geotechnical and Underground Engineering,\\ Shandong University, Jinan 250061, China}}

\begin{abstract}
{Establishing a realistic three-dimensional (3D) microstructure is a crucial step for studying microstructure development of hardened cement pastes. However, acquiring 3D microstructural images for cement often involves high costs and quality compromises. This paper proposes a generative adversarial networks-based method for generating 3D microstructures from a single two-dimensional (2D) image, capable of producing high-quality and realistic 3D images at low cost. In the method, a framework (CEM3DMG) is designed to synthesize 3D images by learning microstructural information from a 2D cross-sectional image. Experimental results show that CEM3DMG can generate realistic 3D images of large size. Visual observation confirms that the generated 3D images exhibit similar microstructural features to the 2D images, including similar pore distribution and particle morphology. Furthermore, quantitative analysis reveals that reconstructed 3D microstructures closely match the real 2D microstructure in terms of gray level histogram, phase proportions, and pore size distribution.
The source code for CEM3DMG is available in the GitHub repository at: https://github.com/NBICLAB/CEM3DMG.
}
\end{abstract}

\begin{keyword}
\texttt{Hardened Cement Paste}\sep
\texttt{3D Microstructure}\sep 
\texttt{Generative Adversarial Networks}

\end{keyword}

\end{frontmatter}

\section{Introduction}
Cement, a fundamental component of construction materials, is ubiquitously utilized in the building industry. The cement hydration process accompanies microstructure development, directly impacting the performance of cement \cite{microstrength,microdurability}. Therefore, investigating the microstructure development during the hydration process has long been a popular topic in cement materials research \cite{cementhydration1,cementhydration2}. Over the past few decades, computational models have emerged as a cost-effective and efficient means to simulate cement microstructure development and investigate hydration behavior \cite{HYMOSTRUC,bentz2000cemhyd3d}. Within computational models of microstructure development, establishing an initial cement microstructure (typically in three-dimensional (3D) space) is a crucial step, and realistic microstructures are essential for ensuring precision in simulations.

\begin{figure*}
 \centering	\includegraphics[width=0.90\textwidth]{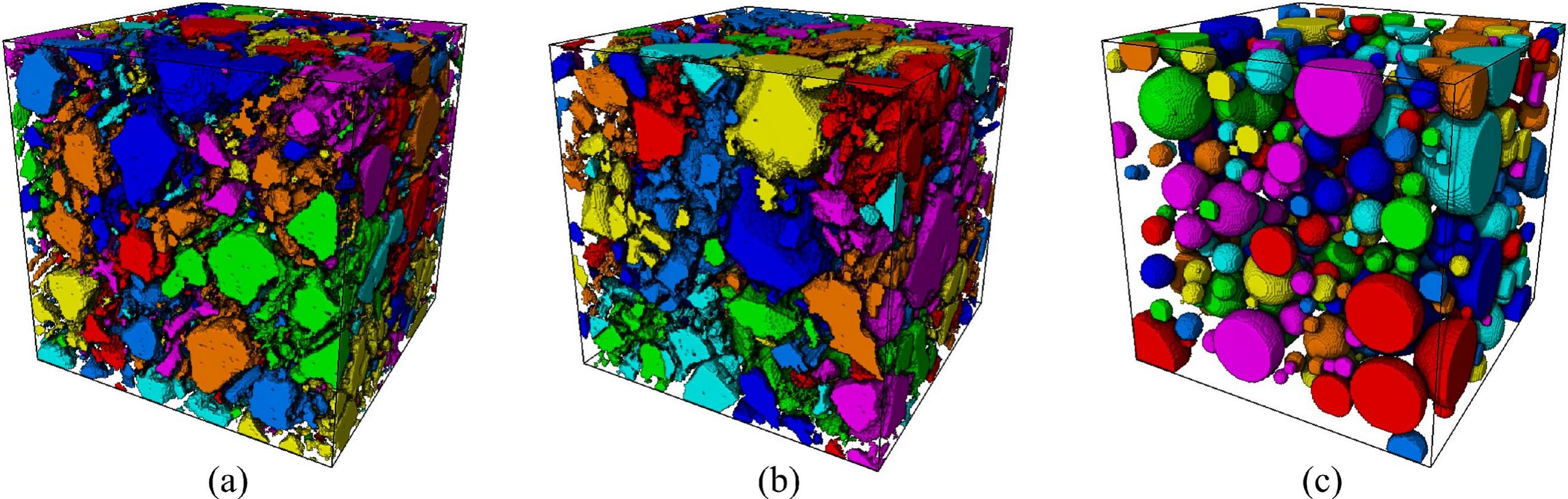}
		\caption{Various 3D microstructures. (a) represents the microstructure generated by our method, (b) denotes the real microstructure, and (c) indicates the spherical-based microstructure.}
		\label{yuan}
\end{figure*}

In the study of cement microstructure development, various models have been proposed, which can be broadly classified into vector \cite{HYMOSTRUC,jennings1986,uic} and discretization \cite{bentz2000cemhyd3d,HydratiCA} models. To streamline computations, these models often assume that cement particles are spherical (see Figure \ref{yuan} (c)), which oversimplifies real particle geometries and may deviate from actual conditions. Bullard et al. recognized this issue related to particle shape and investigated its influence on the process of simulating Portland cement hydration \cite{bullard-particleshape}. The study indicated a significant impact of real particle shapes on the properties of cement. Despite efforts to enhance the realism of microstructures, such as introducing polyhedron and irregular shapes \cite{irregular-shaped,polyhedron}, these approaches have encountered challenges in closely approximating real 3D microstructures.

To obtain real 3D microstructures, direct imaging through scanning instruments is one intuitive approach. Microtomography (Micro-CT), as a non-destructive 3D imaging device, has gained popularity in cement-based materials research \cite{CTapp111,CTapp333}. However, Micro-CT is limited by imaging quality and struggles to provide high-resolution images for large-size samples \cite{CTlargeno}. Scanning Electron Microscopy (SEM) provides high-quality microstructural images, but is limited to capturing only two-dimensional data, which makes it hard to obtain 3D images with rich spatial information. Although Focused Ion Beam Scanning Electron Microscopy (FIB-SEM) \cite{1998FIBSEM} has improved SEM to enable 3D imaging \cite{FIBapp1,FIBapp2}, it is limited by scanning range and sample destruction. Additionally, acquiring microstructures through scanning cement samples for each simulation is costly and impractical. Therefore, a low-cost method to generate high-quality, realistic 3D microstructures is needed to address current limitations.

Compared to 3D microstructural images, high-quality 2D images can be obtained at a lower cost. From the perspective of image analysis, microstructural images of cement generally exhibit isotropy. It is feasible to reveal certain 3D structural characteristics by processing the 2D image \cite{bentz1997three}. With the increased efficiency and accessibility of computers capturing scientists’ attention, generating 3D microstructural images from real 2D exemplars via computational methods has emerged as a feasible alternative.

At the early stage, 3D microstructural image generation efforts focused on generating specific phases, such as pores and particles, from the 2D image of cement \cite{bentz1997three, qiu20213d, monituihuo}. 
As a representative example, Bentz introduced a technique \cite{bentz1997three} that utilizes stereology adjustments and the autocorrelation function to generate 3D cement microstructural images from 2D SEM images.
In these methods, segmentation of the initial 2D image into distinct phases is a prerequisite to streamline the intricate microstructure. While the segmentation simplifies the computational process and reduces the difficulty of image generation, this segmentation operation transforms microstructural images into binary or multivalued formats. This transformation compresses color gradation and fine details, leading to the loss of microstructural information. Thus, such methods make it hard to convert 2D microstructures into 3D counterparts realistically.

\begin{figure*}[!h]
	\centering
		\includegraphics[width=0.95\textwidth]{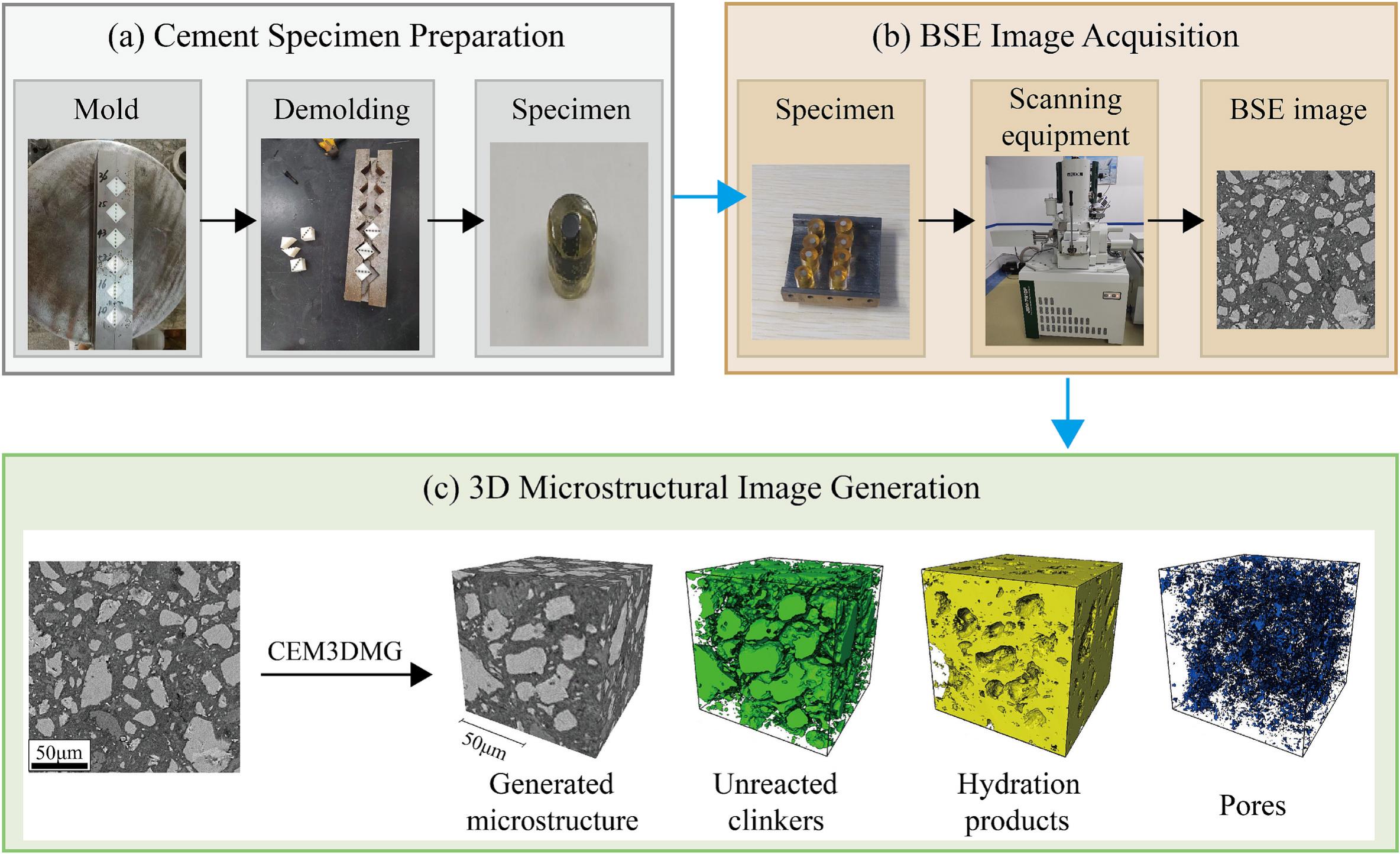}
		\caption{3D microstructural image acquisition process.
		(a) the processed of hardened cement paste specimen preparation, (b) 
        the processed of specimen is scanned to obtain BSE images, 
		(c) the 3D microstructural images are generated from a given single 2D BSE image. 
		}
		\label{total1}
\end{figure*}

To faithfully generate the 3D microstructural image of cement, it is essential to synthesize 3D images from the real 2D microstructural images.
Guided by the principles of solid texture synthesis \cite{perlin1985,peachey1985solid}, Tang et al. employed the Markov Random Field to generate 3D microstructural images from a given 2D Micro-CT image \cite{tang2016three}. Zhang et al. introduced an innovative approach for 3D generation \cite{zhang2019efficient}, integrating random process theory and stochastic reconstruction techniques. 
However, these methodologies are limited by their model complexity and learning capabilities, restricting their capability to capture complex data patterns in the intricate microstructure of cement. This limitation substantially obstructs accurately capturing detailed microstructural information in cement images.

Recently, deep learning has undergone significant advancements, demonstrating potent learning capabilities essential for capturing complex information \cite{NN1, NN2}. Descriptor-based methods and generative adversarial networks (GAN)-based methods have been widely used for 3D microstructure reconstruction. The descriptor-based methods \cite{transferlearningreconstruction,zhao2022three, cnncement2} typically use pre-trained models (usually VGG-19 \cite{VGGICLR}) as a descriptor tool for feature statistics. Since the VGG model is a general model, it is effective in common image synthesis tasks. However, due to the complexity and diversity of cement microstructure, as well as the limited feature expression capability of the VGG-19 model, descriptor-based methods fail to accurately describe the cement microstructures, hindering the generation of realistic 3D microstructures.

Generative adversarial networks have been demonstrated to possess the potential to model arbitrary data distributions. GAN-based methods \cite{GAN-1,GAN-2,GAN-3,GAN-4} have been shown to be very capable in 3D microstructure reconstruction. However, these methods rely on 3D images as training data. In addition to that, Kench et al. proposed a GAN-based method (SliceGAN) for 3D reconstruction based on 2D images \cite{slicegan}, and built a library (MicroLib) \cite{microlib} of 3D microstructures generated from 2D micrographs. Although SliceGAN displays outstanding performance on binary images with simplified structures, the reconstruction quality of SliceGAN deteriorates with increasing microstructure complexity. Moreover, SliceGAN does not account for the multi-scale characteristics of microstructures, which limits its capability to generate realistic 3D microstructures of cement materials. In addition, although 3D reconstruction techniques based on 2D images have attracted attention in material science, these techniques have not been fully explored in extremely complex cement microstructures.

In this paper, we design an improved GAN-based 3D microstructure generation method from a 2D cross-sectional image, which can capture complex and multi-scale cement microstructure information and synthesize realistic 3D microstructural images. This method includes three steps: the specimen preparation, the 2D image acquisition, and the 3D microstructural image generation. In the microstructure generation process, a \textbf{3D} \textbf{M}icrostructural image \textbf{G}eneration framework of \textbf{CEM}ent (CEM3DMG) is proposed. Given the complexity and multi-scale nature of the cement microstructures, CEM3DMG adopts the concept of GAN \cite{ijcai2023p196, NIPSGAN} to learn the arbitrary data distribution (microstructure information) within cement images. Moreover, a multi-scale learning strategy allows CEM3DMG to capture microstructure information at various scales. Additionally, we adopt a block-wise synthesis strategy, enabling CEM3DMG to “block-by-block” generate large-size 3D microstructural images. In the experiments, the generated 3D images and the real 2D images are qualitatively analyzed through visual observation. Subsequently, the generated 3D images are quantitatively compared with the real 2D BSE images in terms of gray level histogram, various phases proportions, and pore size distribution (PSD). The effectiveness and stability of the proposed method are further verified by quantitatively and qualitatively analyzing the similarity between real 3D images and generated 3D images.

\begin{table}[]
\renewcommand{\arraystretch}{1.0}
\centering
\caption{\textcolor{black}{Chemical compositions of cement sample.}}
\label{compositions}
\begin{tabular}{cccc}
\hline
Type        & Sample A    & Sample B & Sample C     \\ \hline
C3S (\%)   & 66.96 & 64.92 & 61.48  \\
C2S (\%)   & 8.7  & 8.15 & 14.94  \\
C3A (\%)   & 6.21  &  7.23 & 5.24  \\
C4AF (\%)  & 11.3  & 9.61  & 10.05 \\ \hline
Age (day)       & 18    & 11  & 23      \\
W/C       & 0.3   & 0.3  & 0.35    \\ \hline
\end{tabular}
\end{table}

\section{Experimental Method}
\label{method}
In this section, Figure \ref{total1} illustrates the experimental methodology, comprising three distinct stages. Initially, specimens of hardened cement paste are prepared (section \ref{sectionpreparation}). Subsequently, SEM is utilized to acquire high-quality 2D BSE images of the hardened cement paste (section \ref{sectionimageacquisition}). Finally, realistic 3D microstructural images are generated using the proposed 3D generation framework (section \ref{sectionsynthesis}). 
{Additionally, the experimental configuration (section \ref{configuration}) and the analysis metrics (section \ref{criteria}) are described.}

\subsection{Specimen Preparation}
\label{sectionpreparation}
In our experiments, we prepare three types of cement samples. The main chemical composition of these samples is determined using X-ray fluorescence (XRF). Table \ref{compositions} includes details on the main chemical components, hydration age, and water-cement ratio of the three samples. The proportions of the four main chemical components (C3S, C2S, C3A, and C4AF) are estimated using the Bogue formula \cite{bogue1955}.

The preparation process for these cement specimens is depicted in Figure \ref{total1}(a).
Initially, the cement paste is introduced into a mold, where it is subjected to stirring and vibration to eliminate bubbles. The cylindrical specimens are cured for 24 hours in 95–100\% RH (relative humidity) at $\sim$23 $^{\circ}$C to set. Subsequently, the specimens are demolded and placed in water for curing until reaching the specified age. Following this, the specimens are immersed in anhydrous ethanol to effectively halt the hydration process by removing moisture. Finally, the specimens are dried and impregnated with epoxy resin, solidifying the internal microstructure.

\begin{figure*}[!h]
	\centering
		\includegraphics[width=0.95\textwidth]{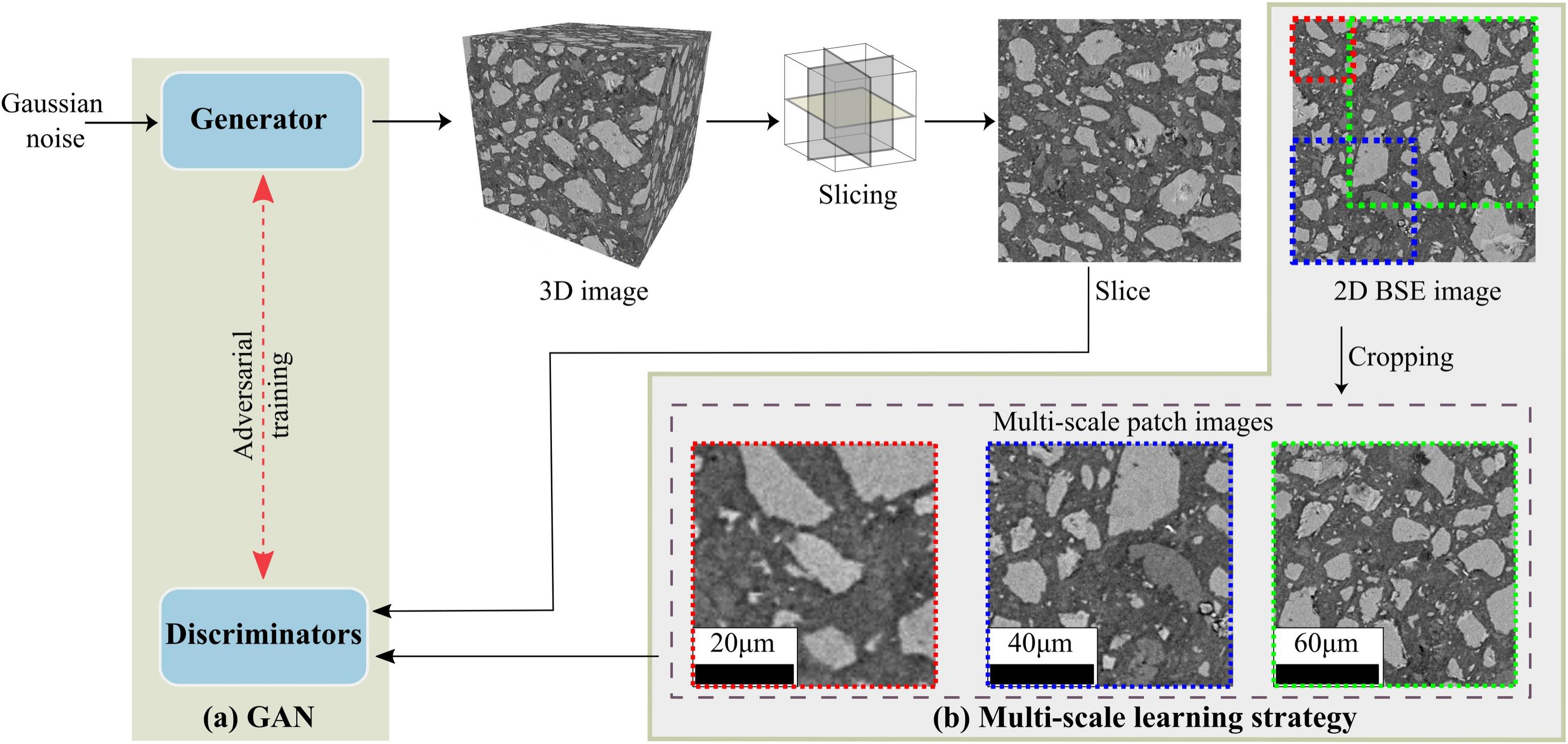}
		\caption{3D microstructural image generation framework (CEM3DMG). (a) the generator and discriminators in GAN-based adversarial training. (b) the multi-scale learning strategy.
		}
		\label{model}
\end{figure*}

\subsection{BSE Image Acquisition} 
\label{sectionimageacquisition}
The process for obtaining BSE images is shown in Figure \ref{total1}(b). 
Initially, the specimens are ground and polished using the automatic grinding and polishing machine with metallographic sandpaper, cloth, and polishing powder. Following grinding and polishing, the specimen surface is cleaned using an ultrasonic cleaner, exhibiting a mirrorlike reflective surface. This surface facilitates the acquisition of high-quality BSE images. Secondly, as the cement specimen is non-conductive, we apply a Pt coating to the specimen’s surface to prevent image distortion caused by charge accumulation.

The processed specimen is then mounted on the sample table and placed into the scanning equipment. The BSE images are acquired using the backscattered electrons mode of field emission scanning electron microscope (JSM-7610F). Moreover, the specimens are scanned under an accelerating voltage of 15 kV and at a magnification of 200$\times$.

\subsection{3D Microstructural Image Generation}
\label{sectionsynthesis}
As shown in Figure \ref{total1}(c), the 3D microstructural image generation phase begins with a single 2D BSE image. CEM3DMG extends the microstructural characteristics of the BSE image into 3D space, generating realistic 3D microstructural images. Finally, the 3D microstructures of the cement are displayed in various visualization forms.

\subsubsection{CEM3DMG}
CEM3DMG employs the concept of GANs, which have been successfully applied in solid texture synthesis \cite{ijcai2023p196, gramgan}. Moreover, CEM3DMG can generate realistic 3D microstructural images from 2D cement images, owing to the capability of GANs to learn any data distribution.
Figure \ref{model} shows the proposed CEM3DMG, which consists of the 3D microstructural image generator $G$ and the discriminator $D$. 
The generator's role is to create 3D microstructural images, while the discriminator's role is to assess the proximity of the generated 3D images to the data distribution of 2D real images and compel the generator to generate realistic 3D images. 
When the discriminator cannot distinguish between real and generated images, the generator is successfully developed and can be utilized to generate realistic images.

A slicing strategy is incorporated into the CEM3DMG to facilitate the cross-dimensional correlation between 2D and 3D images. 
During adversarial training,  when randomly selected slices of the 3D image closely resemble the provided 2D image, the resulting 3D image approximates the data distribution of the given 2D image, thereby generating realistic 3D images.

Within the CEM3DMG, Gaussian noise $z$ is initially input into the generator $G$ to obtain a 3D image $I$.
\begin{equation}
I = G(z),
\label{3d}
\end{equation}

Subsequently, 2D slices are randomly sampled from the generated 3D image in an orthogonal manner. The discriminators assess these slices against the patches cropped from the 2D BSE image.
Through training (see section \ref{Adversarial Training}), the generator $G$ is gradually optimized to generate realistic 3D images.

\begin{figure}[!b]
	\centering
    \includegraphics[width=0.95\columnwidth]{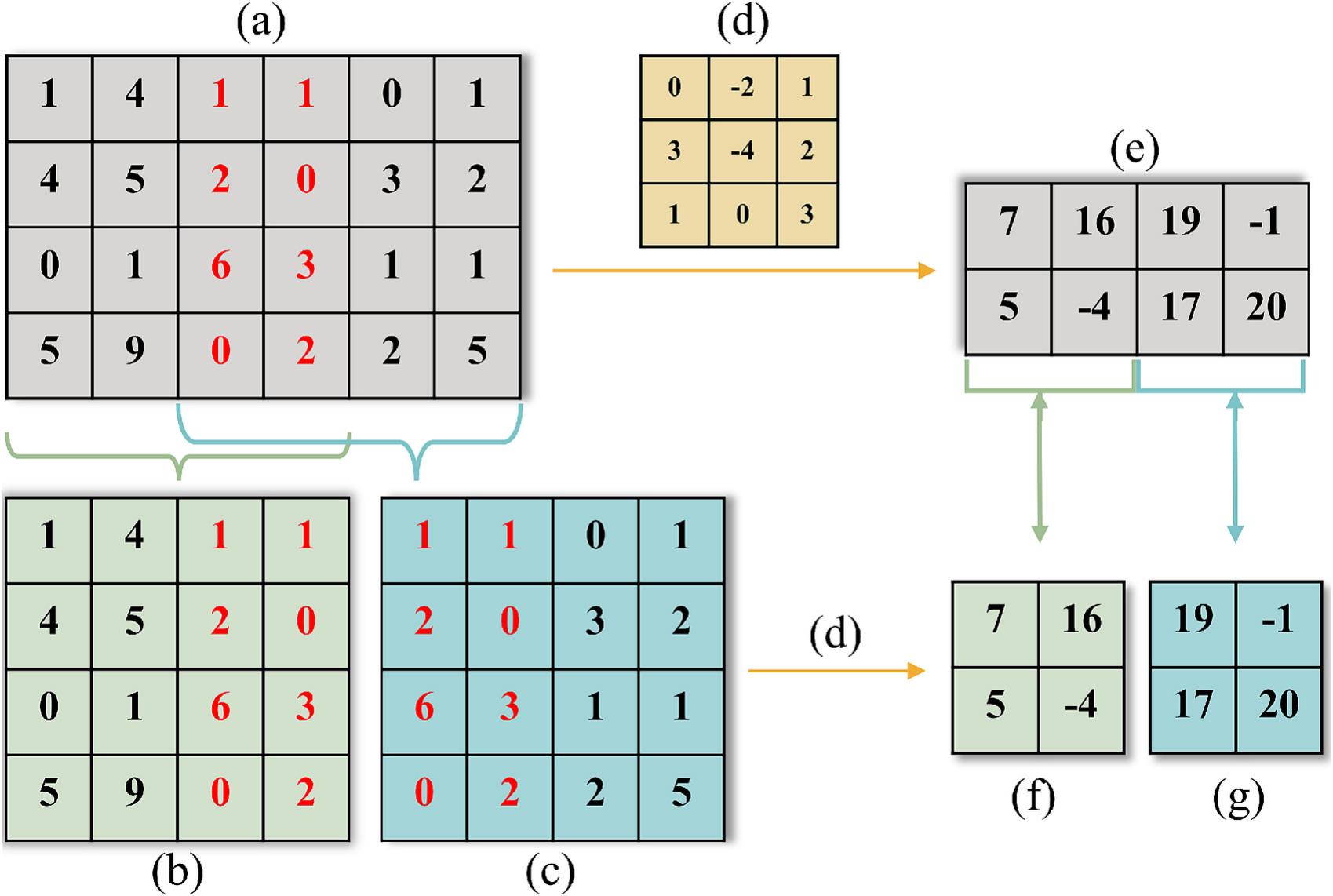}
		\caption{\textcolor{black}{Conceptual diagram of the block-wise synthesis strategy. (a) Input data, (b) and (c) show (a) divided into two smaller input data blocks. The red numbers indicate the overlapping parts between (b) and (c). (d) represents the convolutional operator in the generator (no padding). After processing by the same convolutional operator  
        (d), (a), (b), and (c) produce the output (e), (f), and (g), respectively.
		}}
		\label{block}
\end{figure}

\begin{figure*}
	\centering
		\includegraphics[width=0.96\textwidth]{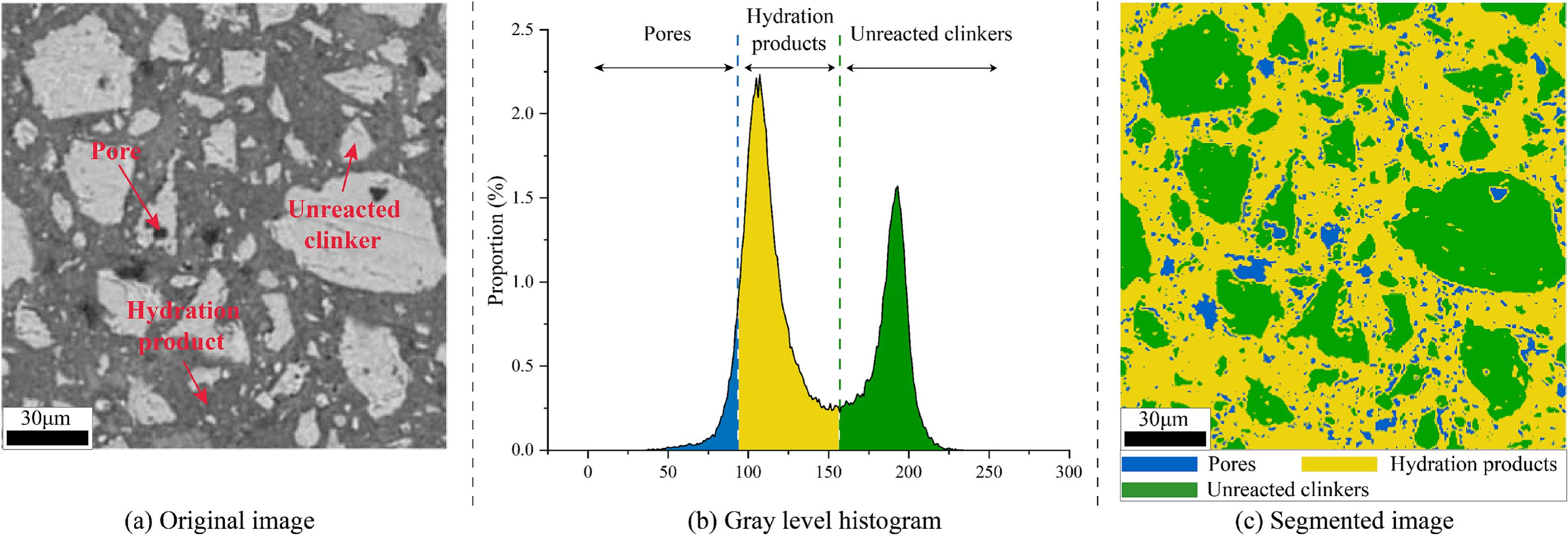}
		\caption{Phases segmentation. Pores, hydration products and unreacted clinkers are segmented based on histogram. In (b), the vertical axis represents the proportion of the number of pixels with the same gray level value to the total number of pixels.}
		\label{seg}
\end{figure*}

\subsubsection{Multi-scale Learning Strategy}
\label{Multi-scale learning}

Given the multi-scale nature of the cement microstructures, a multi-scale learning strategy is employed in CEM3DMG to enable the generator model to learn the data distribution of BSE images at different scales. Figure \ref{model}(b) illustrates the process of obtaining multi-scale images from a single 2D BSE image. Initially, patches of various sizes are randomly cropped from the scanned BSE image. Subsequently, they are rescaled to the same resolution, resulting in images at different scales, where the physical dimensions represented by each pixel vary accordingly. Finally, the generator and discriminator can learn microstructural information from BSE images at different scales, generating realistic 3D microstructural images. In addition, multiple discriminators are employed to learn microstructural information at different scales to alleviate the challenge of a single discriminator learning images at various scales.

\subsubsection{Block-wise Synthesis Strategy}
\label{Block-wise synthesis strategy}
Modeling the 3D microstructure of cement-based materials usually requires multi-size volumetric data \cite{bentzmulti}. Given the limited computational capacity (e.g., memory or GPU memory size), generating large-size 3D images all at once is challenging. In CEM3DMG, the generator is designed with a fully convolutional architecture without padding, allowing the input and output sizes to be flexible (i.e., not constrained to a fixed size). Therefore, we employ a block-wise synthesis strategy, where smaller 3D images are generated first and then pieced together. Figure \ref{block} illustrates the concept of the block-wise synthesis strategy, where large-size outputs are obtained by piecing together small blocks of data. Specifically, during computation, each Gaussian noise input into the generator $G$ overlaps with the neighboring noise parts, ensuring that large-size images are generated while maintaining structure continuity.

\subsubsection{Adversarial Training}
\label{Adversarial Training}
As a technique characterized by stable adversarial training, the idea of WGAN-GP \cite{WGAN-GP} is adopted in CEM3DMG.
Specifically, the generator is optimized based on two components: the discriminator loss and perceptual loss \cite{20CNN, gramgan, gatys2015texture}. The discriminator loss promotes the generator’s learning of the real image data distribution, while the perceptual loss assists the generator in learning the texture style of the image. Together, these components facilitate the generation of realistic microstructural images.
During generator training, the loss function is minimized using the equation:
\begin{equation} 
 \mathcal{L}_G= -\alpha \mathbb{E}[D(x)] + \beta Per(x, u),
 \label{Lossg}
\end{equation}
where $x$ is a 2D slice taken at random from the 3D image generated by $G$, $Per$ refers to the perceptual distance loss between $x$ and cropped image $u$, $\alpha$ and $\beta$ are constants.
Meanwhile, discriminator $D$ is optimized by minimizing its loss function:
\begin{equation}\begin{split}\label{eq2}
 \mathcal{L}_{D}&=\mathbb{E}[D(x)]-\mathbb{E}[D(u)]\\&+\lambda\mathbb{E}[(\parallel\nabla_{r} D(r)\parallel_2-1)^{2}],
 \end{split}
\end{equation}
where $\lambda$ is a constant, and $r$ is a data point uniformly sampled  between $u$ and $x$.

\subsection{Experimental Configuration}
\label{configuration}
PyTorch, a framework for building deep learning models, is utilized to implement the proposed CEM3DMG.
CEM3DMG operates on a system outfitted with an Intel Xeon Silver 4214R Processor and is further bolstered by the capabilities of an NVIDIA A100 Tensor Core GPU. This setup ensures efficient processing and robust computational support for advanced modeling tasks.
In addition, the data and images are analyzed using the software Origin, ImageJ, and Avizo.

\begin{figure}
 \centering	\includegraphics[width=1\columnwidth]{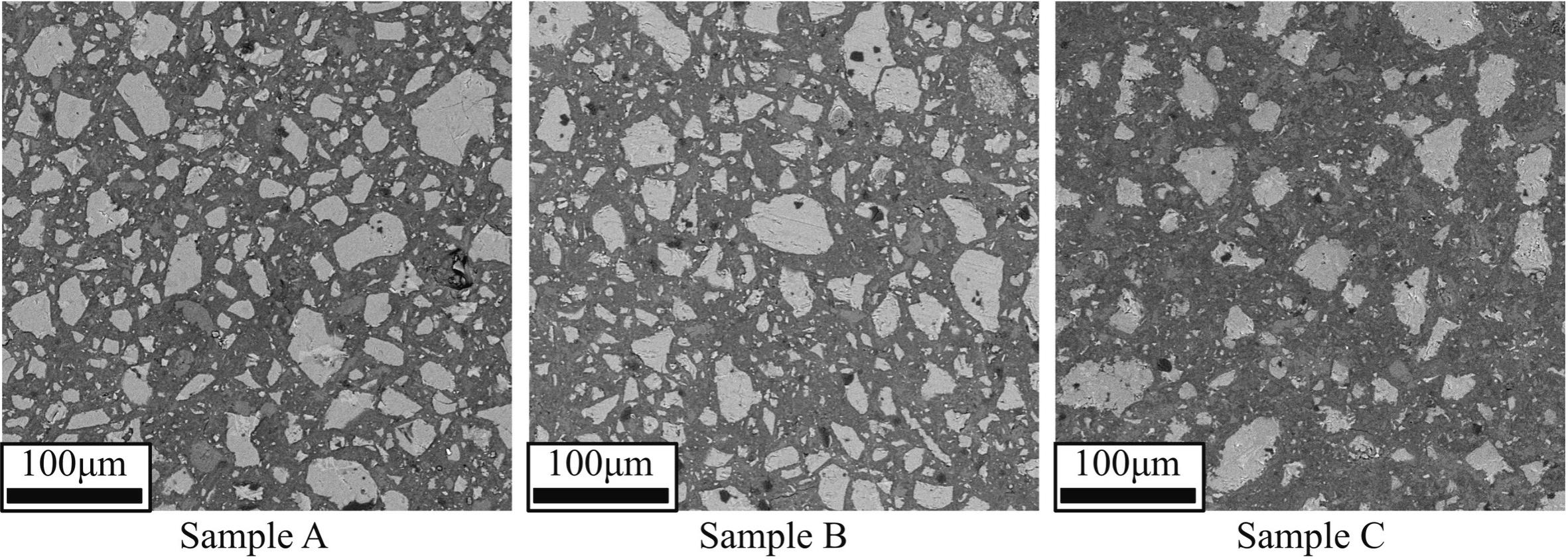}
		\caption{\textcolor{black}{Acquired 2D BSE images of three different types of cement.}}
		\label{sampleimg}
\end{figure}

\subsection{Analytical Evaluation Criteria}
\label{criteria}
The differences between real BSE images and the generated 3D microstructural images are compared to evaluate and analyze the proposed CEM3DMG. During our experimental assessments, various criteria, including gray level histogram, various phases proportions, and pore size distribution, are utilized to analyze the proposed method. Specifically, less discrepancy between these metrics indicates greater similarity in images.

\subsubsection{Gray Level Histogram}
In the microstructure analysis of cement, the gray level histogram is recognized as a classic method \cite{graylevel1,graylevel2}. This histogram illustrates the frequency distribution of each gray level within the microstructure image. The overall trend of the histogram, including the positions of peaks and valleys, provides essential insights into the properties of cement. Therefore, the proximity of the gray level distribution is a crucial factor in assessing the similarity of images. In our experiments, due to differences in image sizes, we standardize the histogram by calculating the proportion of each gray level.

\subsubsection{Various Phases Proportions}
In experiments, the proportions of pores, hydration products and unreacted clinkers are calculated in microstructural images.
These phases proportions, indicative of the material’s density and hydration degree, are key characteristics in the properties of cement \cite{porosityimportant,hydrationdegreeimportant}.
Before calculating various phases proportions, image processing is required, specifically phases segmentation. 
As shown in Figure \ref{seg}, various phases are segmented based on the gray level histogram. Specifically, the overflow method is adopted to find the threshold for segmenting pores \cite{bsewong2006}, and the approach proposed by Scrivener et al. \cite{peakmini} is employed to determine the threshold between hydration products and unreacted clinkers.

\begin{figure*}[!t]
 \centering	\includegraphics[width=0.95\textwidth]{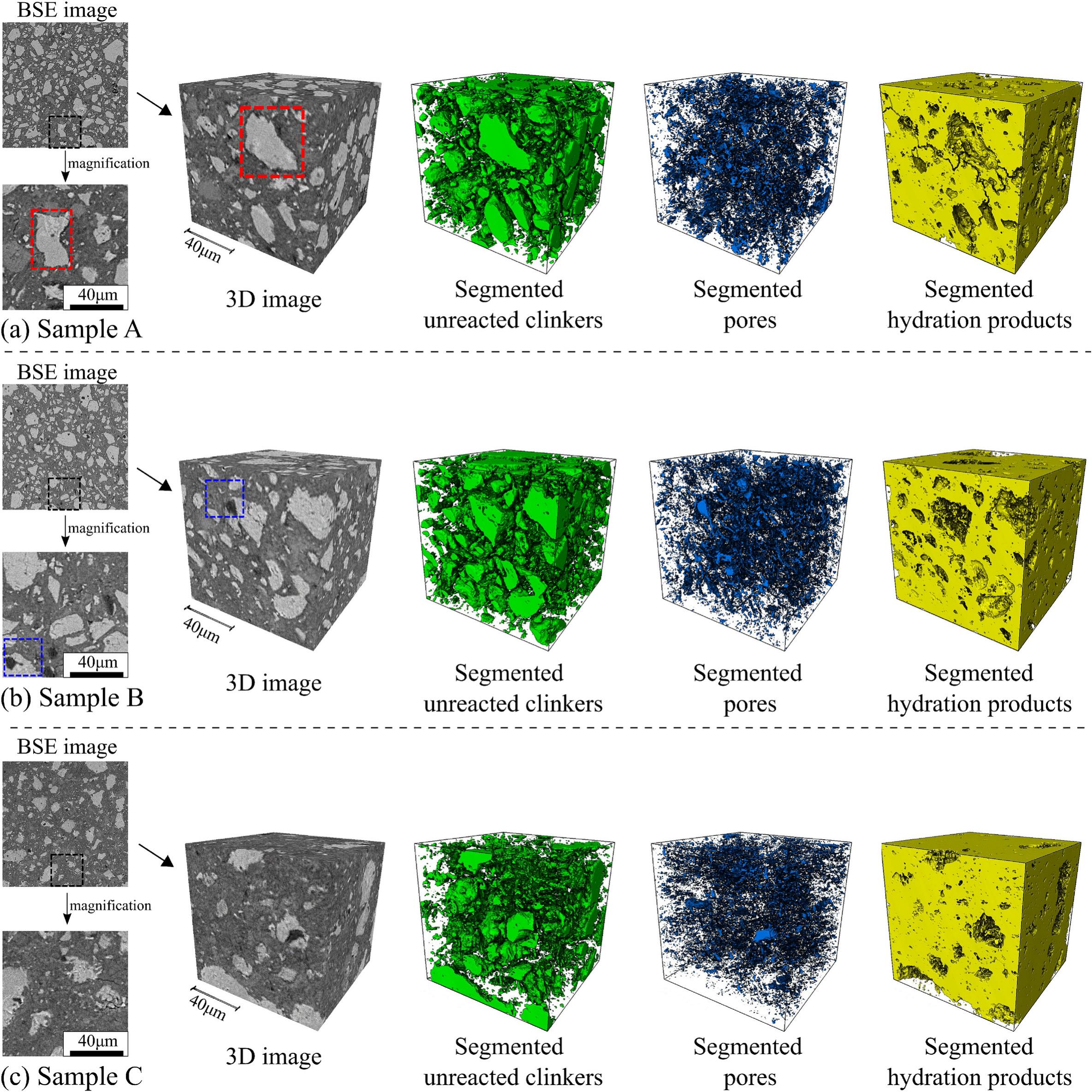}
		\caption{ The generated 3D microstructural images. The left is a real BSE image, and the right are generated microstructural images. Three phases are segmented in microstructure, green denotes unreacted clinkers, blue denotes pores and yellow denotes hydration products. In order to clearly exhibit microstructural details, 3D images with a size of $200 \times 200 \times 200$  pixels$^3$ are displayed in the figure.}
		\label{vision3d}
\end{figure*}

\subsubsection{Pore Size Distribution}
Pores significantly influence several key properties of cement, including its compressive strength and transport properties \cite{porezhongyao1,porezhongyao2}.
Pore size distribution of cement refers to the size range and distribution of pores in cement samples, reflecting its microstructure characteristics\cite{1994poresizedis,2020poresizedis}.
In measuring pore diameters, given that segmented pores typically exhibit irregular shapes, each pore is approximated as either a perfect circle or a perfect sphere in calculations \cite{bselyu2019,perfectsphere1}. For pores in 2D microstructures, the equivalent circular diameter (ECD) of each pore is calculated by equating the area of a perfect circle to the pixel area ($A_p$) of a pore,
\begin{equation}
ECD = \sqrt[2]{4 \times \frac{A_p}{\pi} },
\label{pore2d}
\end{equation}
In the case of 3D microstructures, the equivalent spherical  diameter (ESD) of each pore is calculated by equating the volume of a perfect sphere to the pixel volume of a pore ($V_p$),
\begin{equation}
ESD = \sqrt[3]{6 \times \frac{V_p}{\pi} },
\label{pore3d}
\end{equation}

\section{Results and Discussion}
\label{results and dis}

\subsection{Acquired 2D BSE Images}
In experiments, BSE images are utilized because they can provide phase information, which facilitates us to evaluate the performance of the proposed method through quantitative and qualitative analyses. In addition, BSE images featuring different morphologies are verified in experiments to ensure the effectiveness of CEM3DMG.  Figure \ref{sampleimg} shows three 2D BSE images with different morphologies acquired from hardened cement samples. Each sample is scanned using the SEM, yielding images with a resolution of $800 \times 800$  pixels$^2$ and a pixel size of 0.47 $\mu m$.

\begin{figure*}
 \centering	\includegraphics[width=0.99\textwidth]{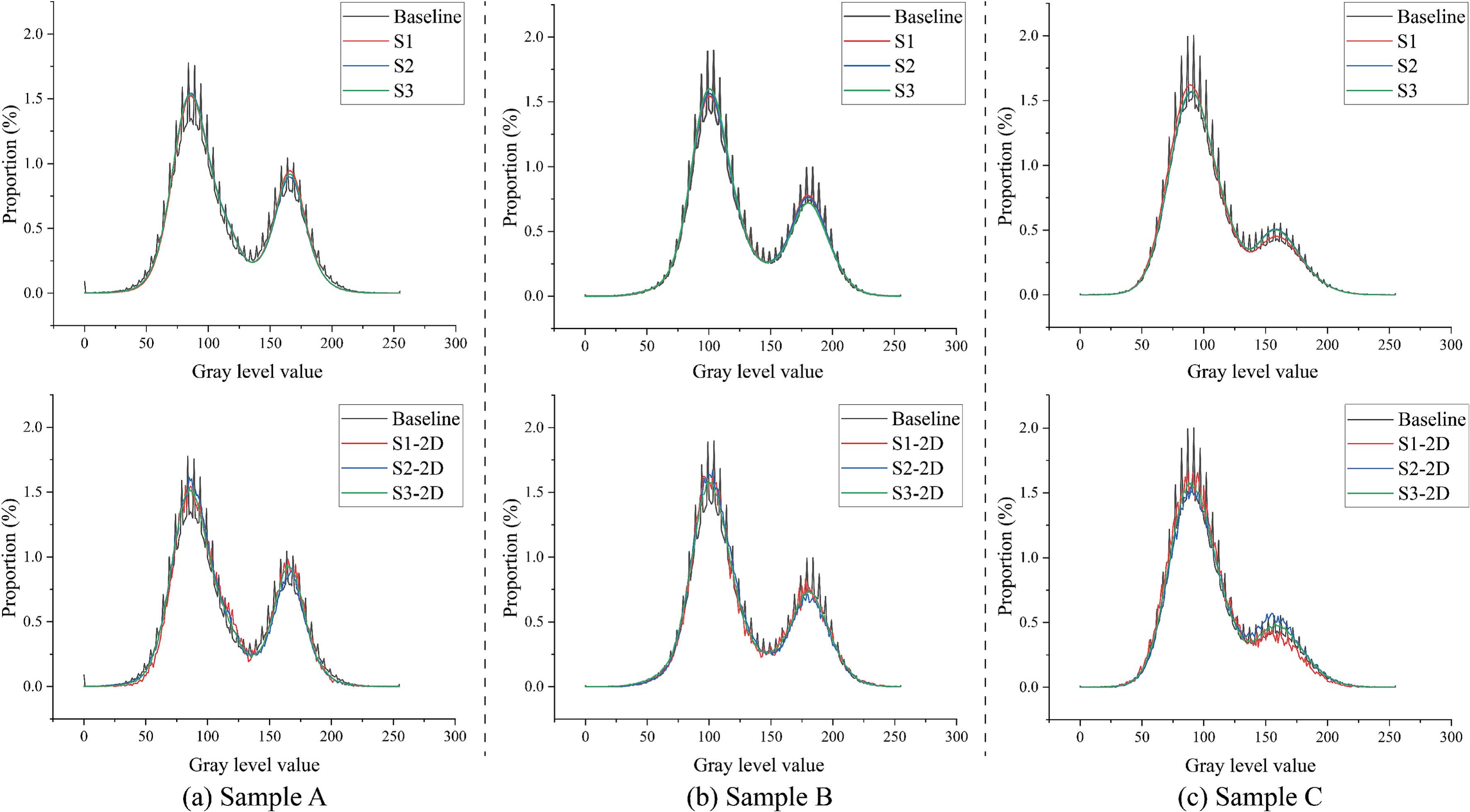}
		\caption{Gray level histograms of 2D and 3D images. The baseline represents the real 2D BSE image, while S1, S2, and S3 denote three 3D images with distinct sizes, respectively. Note that the first row displays the results of the 2D BSE image and the generated 3D image, while the second row shows the results of the 2D BSE image and the 2D slice randomly selected from the generated 3D image.}
		\label{gray3dand2d}
\end{figure*}

\begin{figure*}
	\centering
		\includegraphics[width=0.99\textwidth]{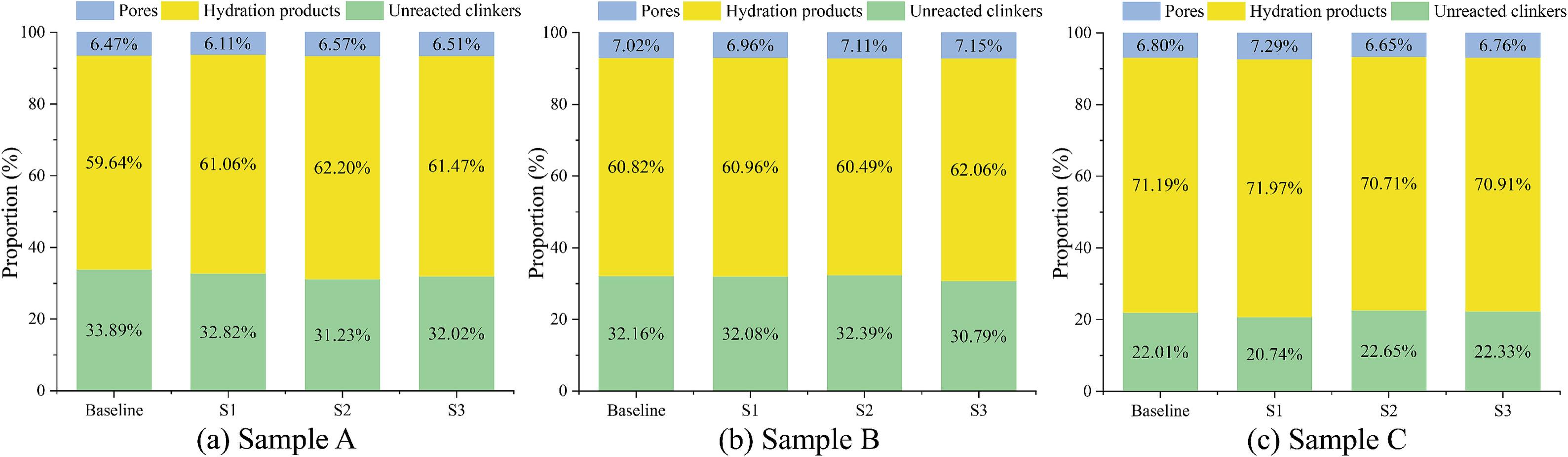}
		\caption{The various phases proportions of generated microstructures and real microstructures. The baseline represents the real 2D microstructure, while S1, S2, and S3 denote three generated 3D microstructures with distinct sizes, respectively.}
		\label{Phases-BSE}
\end{figure*}

\subsection{Qualitative Analysis of 3D Microstructural Image Generation}
This section primarily focuses on visually assessing the generation effectiveness of 3D microstructural images. In this experiment, CEM3DMG generates 3D microstructural images based on 2D BSE images of hardened cement pastes. In Figure \ref{vision3d}, besides the presentation of the 3D microstructure, various phases within the microstructure are also marked with different colors, enabling a clear visualization of the internal structures of the generated 3D microstructures.

As shown in Figure \ref{vision3d}, it is observed that the generated 3D microstructural images visually match the given 2D BSE image. In Figure \ref{vision3d}(a), the cement particles (i.e., unreacted clinkers, as indicated by the red boxes in Sample A, for example) in both the real 2D and generated 3D microstructures exhibit similar morphologies, with various-sized particles reflected in the generated 3D microstructure.

In Figure \ref{vision3d}(b), Sample B exhibits obvious pores and is closely connected to the particles (for instance, with some pores encapsulated by cement particles, as indicated by the blue boxes). Upon closer examination, these pore characteristics are preserved in generated microstructural images, as marked by the blue boxes. As shown in Figure \ref{vision3d}(c), the generated 3D images have similar microstructural characteristics to the given 2D BSE image. Due to the longer hydration time of Sample C, Sample C has a higher proportion of hydration products compared to Sample A and Sample B. A similar phenomenon can be observed in the generated 3D microstructure, where the proportion of hydration products is higher than in Sample A and Sample B.

Therefore, from a visual standpoint, CEM3DMG effectively captures the microstructural information in the given 2D BSE image and generates high-quality and realistic 3D microstructures of cement.

\subsection{Quantitative Analysis of 3D Microstructural Image Generation}
In studies on cement paste microstructures, the size of representative volume element (RVE) and representative surface element (RSE) is typically around 100 $\mu m$ \cite{RVE1,RVE2,RVE3}. In our experiments, to ensure the reliability of validation results, larger sizes for both RVEs and RSEs are adopted.
For the original 2D BSE images, the RSE encompasses the entire image, i.e. $800 \times 800$ pixels$^2$ with 0.47 $\mu m$ per pixel (equivalent to $376 \times 376$ $\mu m^2$).
For the 3D microstructural images generated, three different RVE sizes are evaluated: $200 \times 200 \times 200$ pixels$^3$ (S1, $94 \times 94 \times 94$ $\mu m^3$), $400 \times 400 \times 400$ pixels$^3$ (S2, $188 \times 188 \times 188$ $\mu m^3$), and $800 \times 800 \times 800$ pixels$^3$ (S3, $376 \times 376 \times 376$ $\mu m^3$).

\subsubsection{Gray level histogram}
The gray level histogram represents quantitative information about the gray level distribution of BSE images. Therefore, the histograms of real BSE images and generated 3D images with different sizes (i.e., S1, S2, S3) are first compared and analyzed for three samples. Subsequently, the histograms of slices extracted from 3D images are compared with those of real BSE images.

\begin{figure*}[!h]
 \centering	\includegraphics[width=0.96\textwidth]{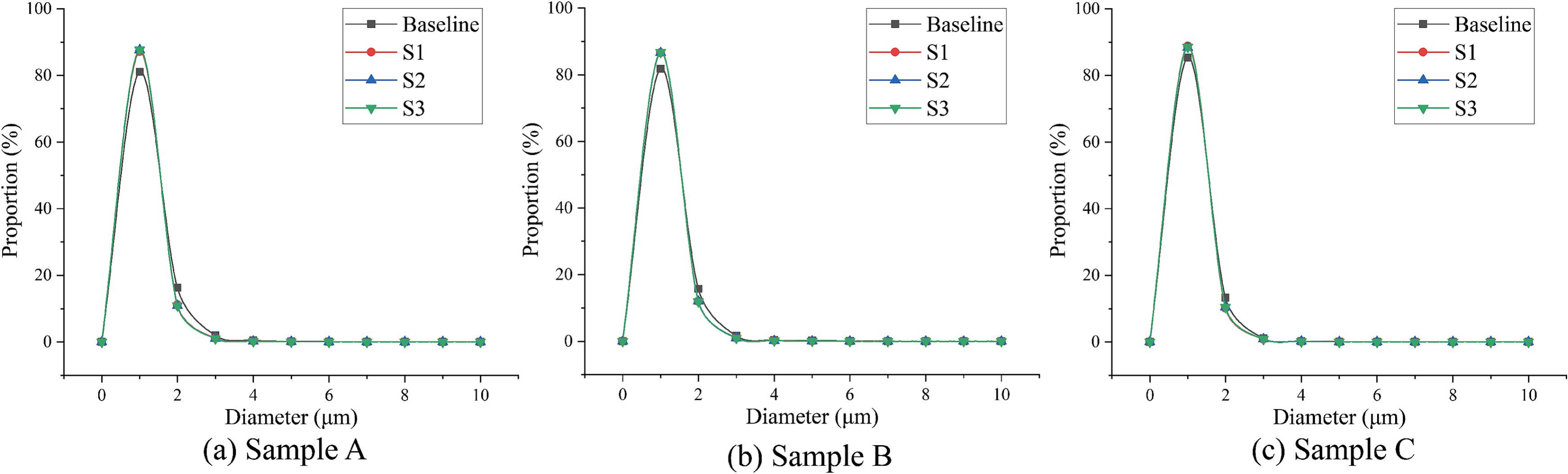}
		\caption{Pore size distribution of 2D microstructure and different-size 3D microstructures. The baseline represents the real 2D microstructure, while S1, S2, and S3 denote three generated 3D microstructures with distinct sizes, respectively.}
		\label{poresizedis}
\end{figure*}

\begin{figure}
	\centering
		\includegraphics[width=0.99\columnwidth]{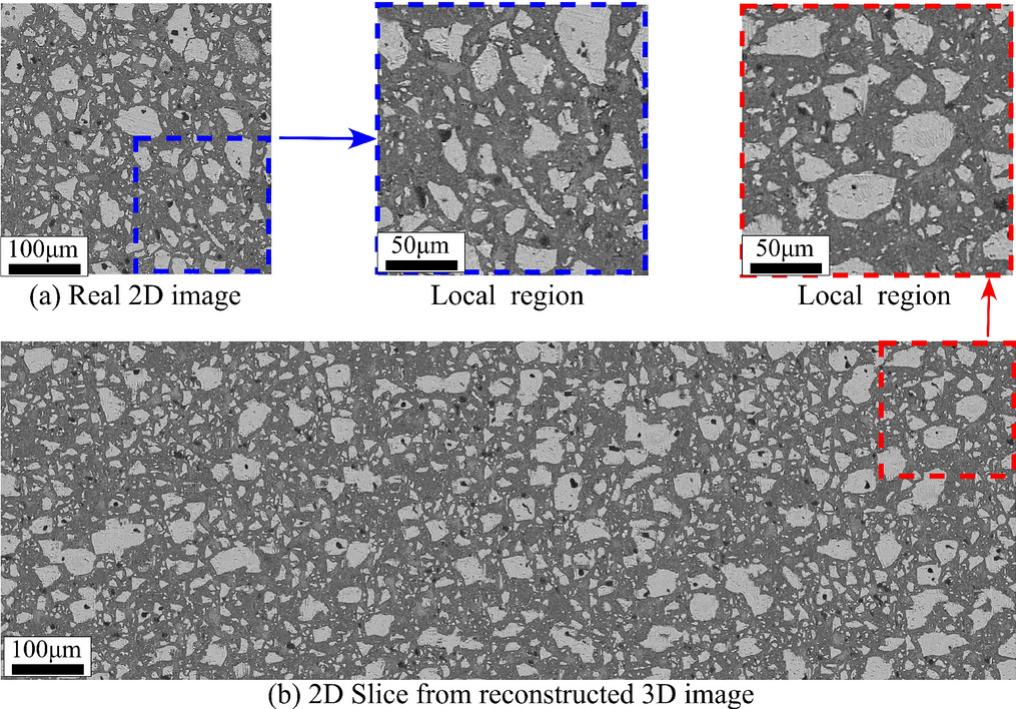}
		\caption{The generated large-size microstructural image. (a) real 2D BSE image, (b) the 2D slice from generated 3D image.
		}
		\label{anysize}
\end{figure}

Figure \ref{gray3dand2d} (first row) presents histograms of 2D BSE images and generated 3D images with varying sizes. Additionally, histograms of 2D slices extracted from different orientations within the 3D image and real BSE images are depicted in the second row of Figure \ref{gray3dand2d}. As illustrated in Figure \ref{gray3dand2d}, the gray level distribution trends of the real BSE images (baseline) from every sample and the generated 3D images are consistent. In particular, crucial information such as the positions of peak and valley in the histograms remains consistent. These details are vital for phase analysis, providing critical insights into the composition and characteristics of cement. This result indicates that CEM3DMG can learn color information from the 2D BSE image and extend it to the 3D image.

In Figure \ref{gray3dand2d}, while the overall trend of gray level distribution is various for all three samples, the trend exhibits localized fluctuations (for instance, the histogram of Sample A shows a spike approximately every ten gray levels). In the experiment results, the generated 3D images smooth out these local variations, suggesting that CEM3DMG is challenging to learn thoroughly from images with the locally unstable gray level distribution. Nevertheless, these differences have a minimal impact on image analysis, as the overall trend of histograms has been captured by CEM3DMG.  Furthermore, mitigating these subtle discrepancies in the gray level distribution can be straightforward, for example, by using image processing techniques to smooth the original images or to sharpen the generated images.

In summary, CEM3DMG effectively captures the microstructure of 2D BSE images and generates corresponding 3D microstructural images. The approach’s effectiveness is demonstrated through the analysis of gray level histograms.

\subsubsection{Various Phases Proportions}
For cement pastes, different phases proportions are critical indicators as they are directly related to the structure and properties of the materials. In the experiment, various phases (pores, hydration products and unreacted clinkers) proportions of real 2D microstructures and generated 3D microstructures are analyzed. Figure \ref{Phases-BSE} presents the results of phase proportion analysis for different microstructures. Taking the pores proportion (porosity) as an example, the porosity of the real 2D microstructure and the generated 3D microstructures for each sample is approximative, with a maximum difference (baseline and S1 of Sample C) not exceeding 0.49 \%.

For hydration products, the maximum difference in Sample A is 2.56\% (S2), in Sample B is 1.24\% (S3), and in Sample C is 0.78\% (S1). Regarding unreacted clinkers, the maximum difference in Sample A is 2.66\% (S2), in Sample B is 1.37\% (S3), and in Sample C is 1.27\% (S1). These results demonstrate that the proportions of hydration products and unreacted clinkers in the generated microstructures closely resemble those in the real 2D microstructure.

The analysis of various phases proportions indicates that the generated 3D images align closely with the real 2D BSE images, providing further evidence that CEM3DMG can generate realistic 3D microstructure conforming to the 2D microstructure information.

\begin{figure*}[!h]
	\centering
		\includegraphics[width=0.95\textwidth]{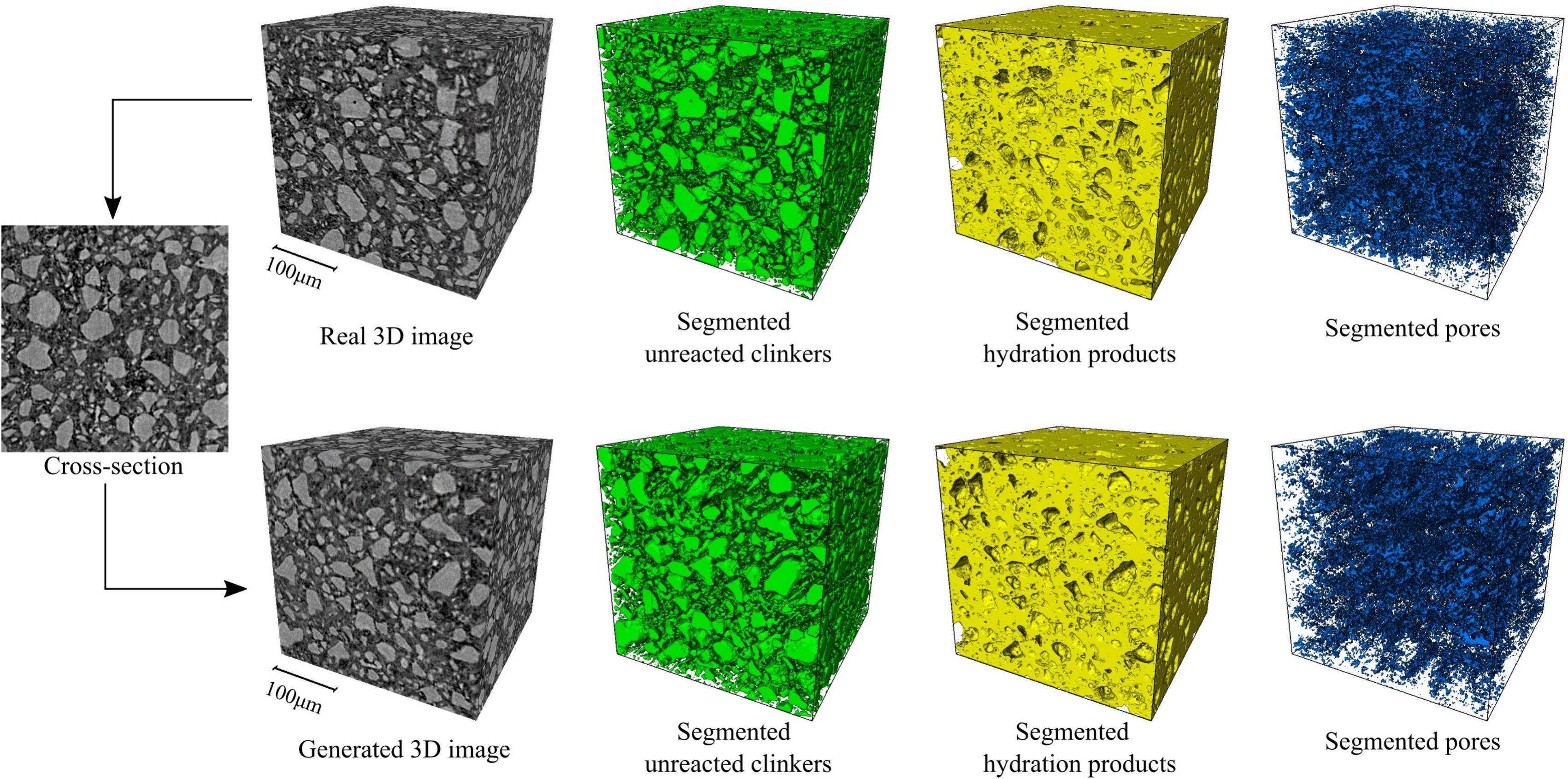}
		\caption{Comparison of real 3D images and generated 3D images. Randomly select a cross-section from a real 3D microstructural image and generate a new microstructure based on the single cross-section.
		}
		\label{CT3D}
\end{figure*}

\begin{figure}[h]
	\centering
		\includegraphics[width=0.95\columnwidth]{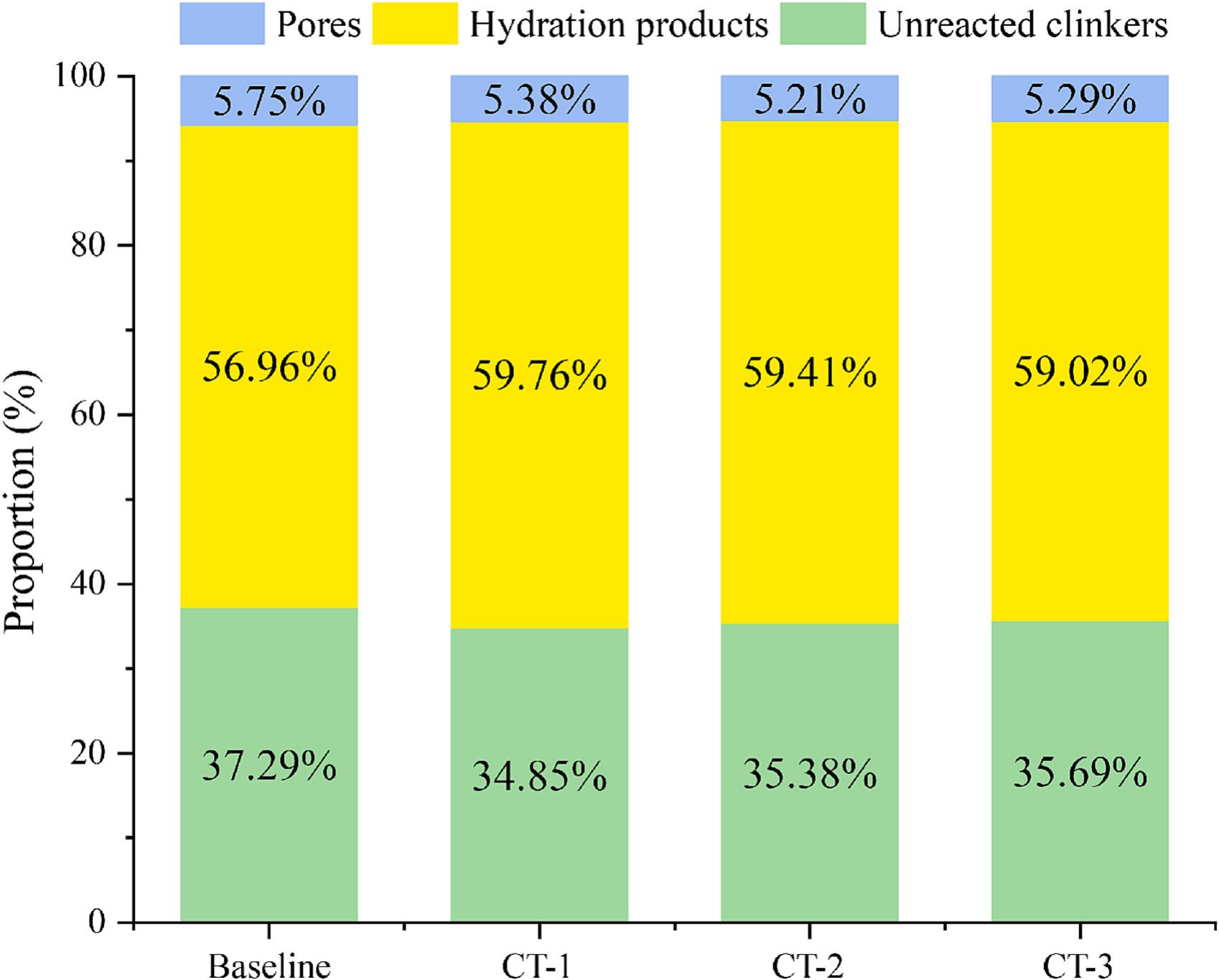}
		\caption{The various phases proportions of generated and real 3D microstructure. The baseline represents the real 3D microstructure, while CT-1, CT-2, and CT-3 denote three generated 3D microstructures, respectively.
		}
		\label{CT_phase}
\end{figure}

\begin{figure}[h]
	\centering
		\includegraphics[width=0.95\columnwidth]{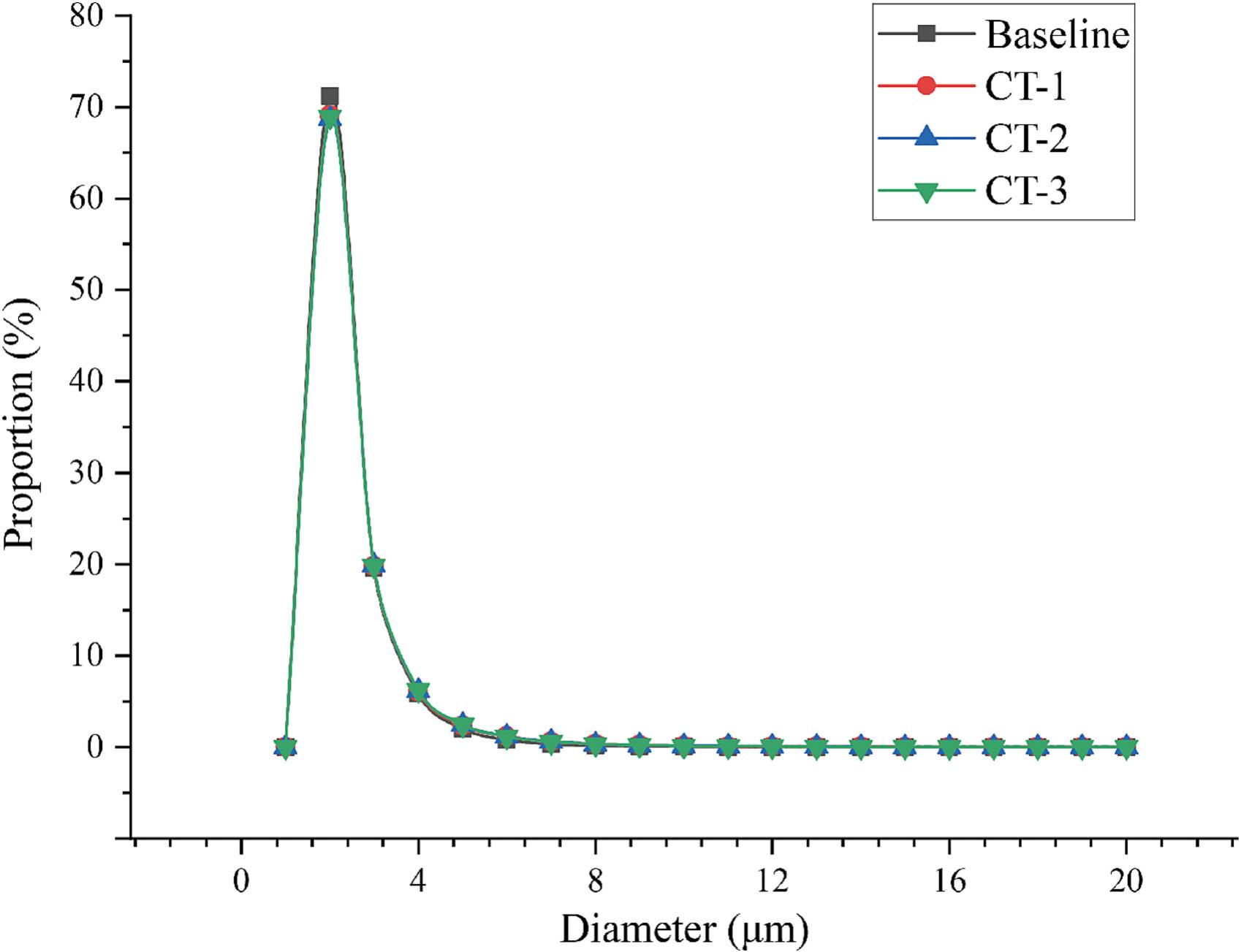}
		\caption{Pore size distribution of real and generated 3D microstructures. The baseline represents the real 3D microstructure, while CT-1, CT-2, and CT-3 denote three generated 3D microstructures, respectively.
		}
		\label{CT3DPSD}
\end{figure}

\begin{figure*}[h]
	\centering
		\includegraphics[width=0.98\textwidth]{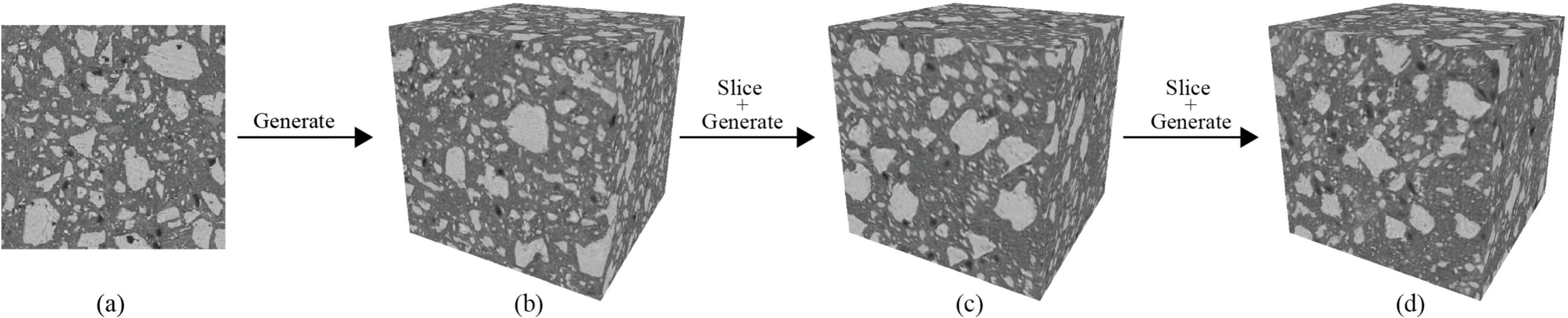}
		\caption{Sequentially reconstruct 3D microstructures based on the generated 2D slice.
		}
		\label{3Dround}
\end{figure*}

\begin{figure}[h]
	\centering
		\includegraphics[width=0.90\columnwidth]{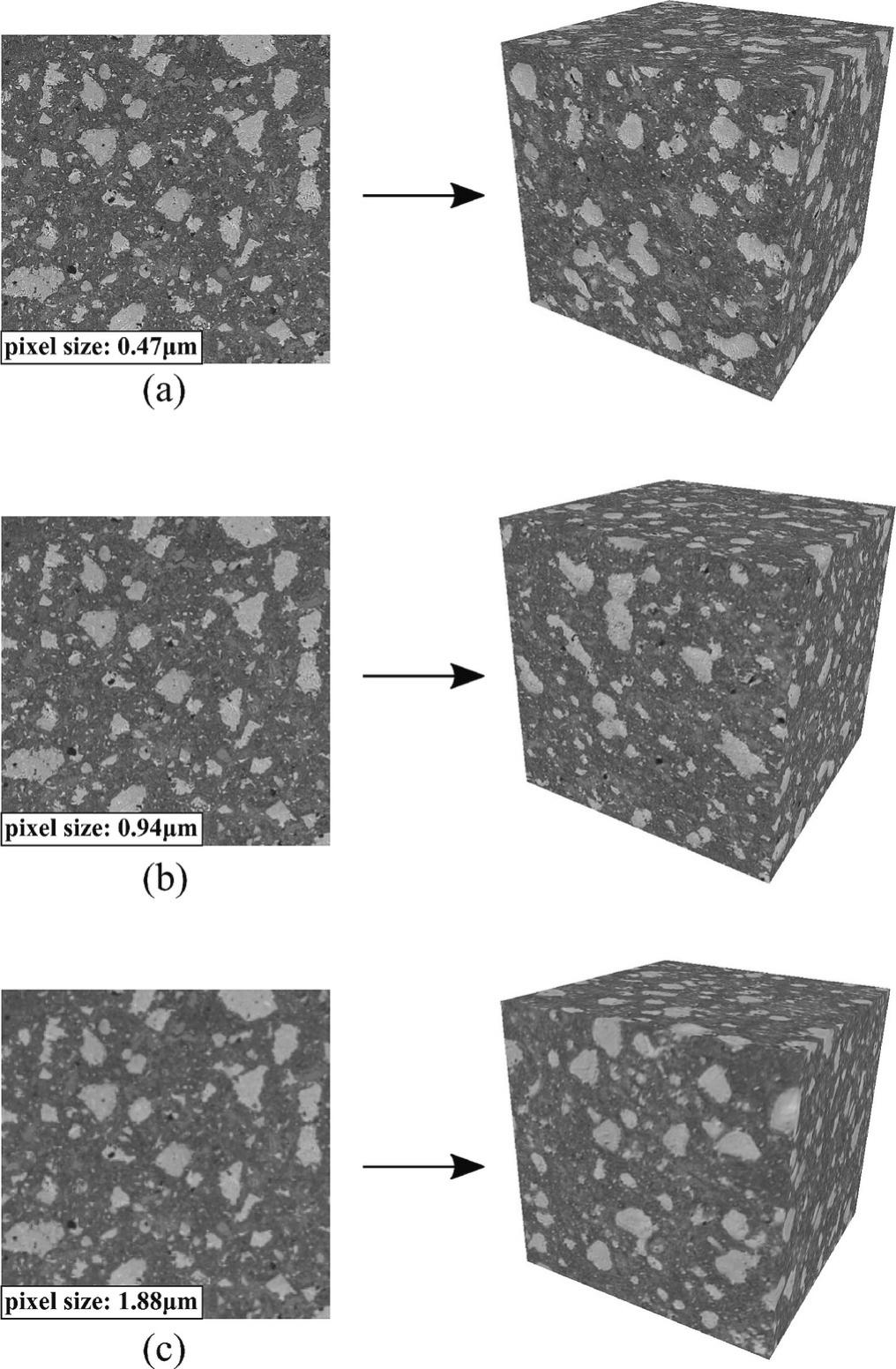}
		\caption{3D images generated from 2D images with different pixel sizes.
		}
		\label{pixel-vis}
\end{figure}

\begin{figure}[h]
	\centering
		\includegraphics[width=0.90\columnwidth]{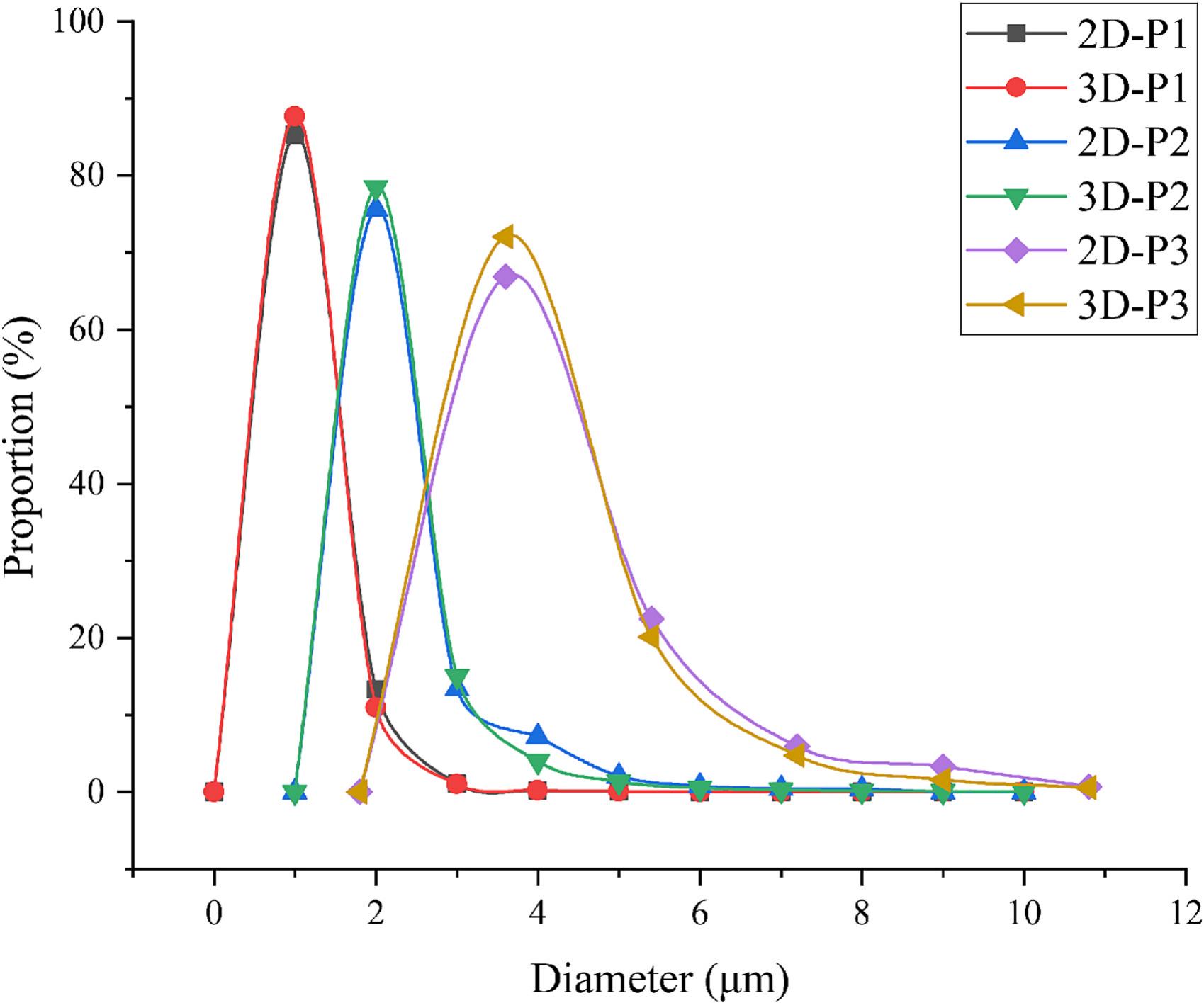}
		\caption{Pore size distribution of 2D microstructure and 3D microstructures. P1, P2, and P3 represent pixel sizes of 0.47$\mu m$, 0.94$\mu m$, and 1.88$\mu m$, respectively.
		}
		\label{pixel-psd}
\end{figure}

\begin{figure}[h]
	\centering
		\includegraphics[width=0.90\columnwidth]{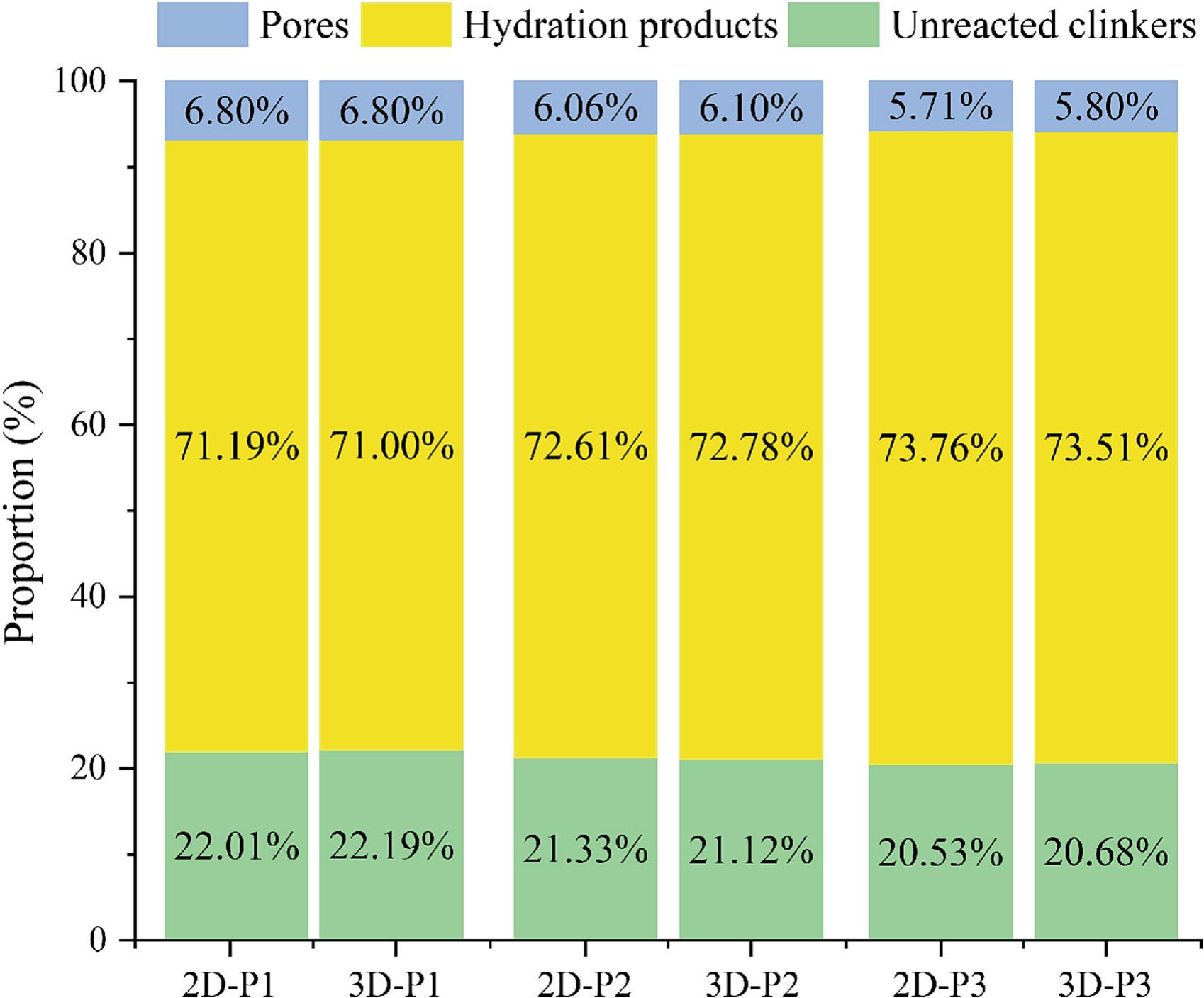}
		\caption{The various phases proportions of generated 3D microstructures and real 2D microstructures. P1, P2, and P3 represent pixel sizes of 0.47$\mu m$, 0.94$\mu m$, and 1.88$\mu m$, respectively.
		}
		\label{pixel-vpp}
\end{figure}

\begin{figure}[h]
	\centering
		\includegraphics[width=0.99\columnwidth]{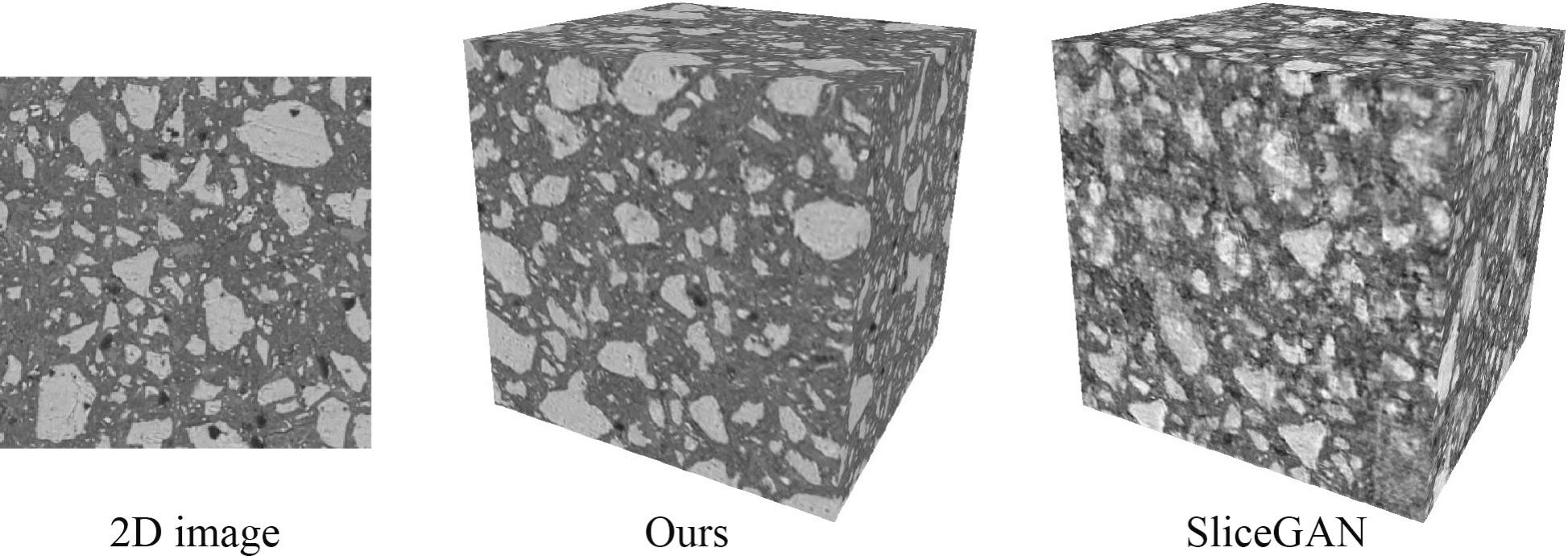}
		\caption{Comparison of reconstruction quality between ours and SliceGAN.}
		\label{duibiimg}
\end{figure}

\subsubsection{Pore Size Distribution}
\label{section PSD}
Analyzing PSD allows for an understanding of the distribution of pores of different sizes within the cement paste, further assisting researchers in comprehending the microstructural characteristics of the material. In the experiment, the PSDs of 2D and variously-sized 3D microstructures are analyzed.

Figure \ref{poresizedis} illustrates the PSD of microstructures, showing that pores with a diameter less than 1 $\mu m$ dominate, suggesting high material density consistent with previously discussed low porosity. Although the PSD of the real 2D and 3D microstructures are overall similar, there are still some variations in tiny pores. In Figure \ref{poresizedis}, it is observed that the most significant discrepancies in pore size occur in pores with diameters less than 2 $\mu m$. As pore size increases, the distribution of pores between the 2D microstructures and 3D microstructures gradually aligns.

The primary reason for this discrepancy can be attributed to the exceedingly small size of these pores, often spanning only a few pixels. Thus, the impact of image noise on CEM3DMG's learning of these tiny pores results in differences between the generated 3D microstructures and real 2D microstructures in tiny pores.

\subsection{Large-size Microstructual Image}
Large-size microstructural images offer more comprehensive microstructural insights for cement research. CEM3DMG employs a block-wise synthesis strategy (section \ref{Block-wise synthesis strategy}) to ensure the image generation at desired size, particularly large-size 3D microstructural images. Figure \ref{anysize} shows the generated large-size microstructural image, where the size of the real 2D image  is $800 \times 800$  pixels$^2$, and the 2D slice of the generated 3D image is $1000 \times 3000$  pixels$^2$. Notably, local details of the same size are displayed for enhanced visual comparison. The results indicate that the characteristics of the real 2D images are retained even in large-size images. Therefore, CEM3DMG can generate 3D microstructural images with large sizes while preserving microstructural characteristics consistent with the given 2D images.

\subsection{Verification on 3D Microstructral Image}
To validate the effectiveness of CEM3DMG, an intuitive solution is to select a 2D slice from a real 3D structure, then use CEM3DMG to generate a new 3D structure from the 2D slice, and compare the generated 3D microstructure with the real 3D microstructure. However, since SEM can only capture 2D microstructural images and fails to obtain real 3D microstructural images, we instead use real 3D micro-CT images for validation in experiments. Additionally, to further assess the stability and effectiveness of CEM3DMG, we also utilize the generated 3D image as ground truth to perform the same validation process.

The 3D micro-CT image data from the Visible Cement Dataset (available from http://visiblecement.nist.gov) is obtained using the European synchrotron radiation facility in Grenoble \cite{bentz2002visible}. In the micro-CT dataset, specimens are scanned at a resolution of 0.95 $\mu m$ per pixel. The size of the micro-CT image is $300 \times 300 \times 300$ pixels$ ^ 3$ ($285 \times 285 \times 285$ $\mu m$$^3$).
In the experiment, a 2D slice is first selected from real 3D Micro-CT images, and then 3D microstructural images are generated based on the 2D slice.

Figure \ref{CT3D} shows that a 3D microstructural image is generated from a single 2D slice taken from the real 3D image. The image size and pixel size of the generated 3D structure are the same as the real 3D structure, i.e., $300 \times 300 \times 300$ pixels$ ^ 3$ with a pixel size of 0.95 $\mu m$. For the Micro-CT images, we adopt the idea of Kim et al. to determine the threshold value of the pores \cite{CTsegment}, and the segmentation value between hydration products and unreacted clinkers using the method of Scrivener et al. \cite{peakmini}. In visual comparison of the 3D images, the internal pores, unreacted clinkers and hydration products are displayed. From a visual perspective, the generated 3D microstructure is similar to the real 3D microstructure, albeit relying only on a single 2D image. The most notable aspect lies in the morphology of cement particles, where the real and generated microstructures closely resemble each other, making it difficult to distinguish them through visual observation.

In addition to visual analysis, as shown in Figure \ref{CT_phase}, the proportions of different phases are illustrated. It can be observed that the proportions of the three phases between the real and generated microstructures are generally consistent. Although the morphology of clinkers and hydration products in the generated microstructure closely resembles that of the real microstructure, the proportions of phases within the microstructure are not entirely consistent. On one hand, this suggests that CEM3DMG still requires further refinement for improved accuracy in learning these microstructural details. On the other hand, relying solely on histogram-based segmentation methods for different phases (hydration products and clinkers) within the microstructure makes it challenging to achieve entirely accurate segmentation. Figure \ref{CT3DPSD} presents a comparison of the PSD between the real and generated 3D microstructures, revealing a close overall distribution of pore sizes. The experiments demonstrate that the PSD in the generated microstructures closely resembles that of the real microstructure.

Figure \ref{3Dround} demonstrates the verification process and results by using the generated 3D images as ground truth. To ensure the representativeness and information richness of the initial 2D microstructural image, the experiment starts from a single 2D BSE image which is sized at $266 \times 266$ pixels$ ^ 2$ with a resolution of 0.94$\mu m$ per pixel (i.e., 250 × 250 $\mu m^2$), as shown in Figure \ref{3Dround}(a). After generating a new 3D microstructure image using the CEM3DMG method, we use this image as the ground truth. Then, the 3D image is sliced, and a slice is used to generate the next generation of 3D structures using CEM3DMG. Similarly, the generated 3D images maintain the same dimensions (266 × 266 × 266 pixels$ ^ 3$) and pixel size. As shown in Figure \ref{3Dround}, the generated 3D images (Figure \ref{3Dround}(b), \ref{3Dround}(c), \ref{3Dround}(d)) maintain similar microstructural characteristics to the original 2D BSE image, confirming the ability of CEM3DMG to consistently capture and extend 2D microstructural details to 3D structures. This result demonstrates that the proposed CEM3DMG can generate realistic 3D microstructural images and is effective and stable.

\subsection{The Influence of Pixel Size on Microstructure Analysis}
In a fixed-size scanning area, a smaller pixel size, which corresponds to higher spatial resolution yields clearer images with more observable microstructural details. In this section, the effect of pixel size on microstructure analysis is discussed. Figure \ref{pixel-vis} shows the 3D microstructural images generated from 2D images with different pixel sizes under the proposed method. The spatial resolutions for each 2D image are $800 \times 800$, $400 \times 400$, and $200 \times 200$, and the pixel sizes are 0.47$\mu m$ (P1), 0.94$\mu m$ (P2), and 1.88$\mu m$ (P3), respectively. Moreover, the resolution and pixel size of the synthesized 3D images are consistent with the corresponding 2D images. It can be observed that the quality of the generated 3D images increases as the pixel size of the 2D images decreases, because 2D images with smaller pixel sizes contain more microstructural details. This phenomenon reflects that 2D images with rich microstructural information can aid the proposed method in generating high-quality 3D images.

In terms of pore size distribution, smaller pore features can be observed in images with a smaller pixel size. As shown in Figure \ref{pixel-psd}, the generated 3D microstructure aligns well with the 2D structure, closely reflecting the similar PSD characteristics. With increased pixel size, tiny pores become indistinguishable and thus unmeasurable due to decreased spatial resolution. Since the pixel size serves as the minimum computational unit, a larger pixel size leads to cumulative error in pore size computation. Conversely, when the pixel size is small, the error in pore size distribution computation is reduced. The results indicate that a smaller pixel size is conducive to more accurate analysis of pore characteristics.

Figure \ref{pixel-vpp} presents the various phases proportions in 2D and 3D microstructures with different pixel sizes. Pixel size, as discussed previously, directly impacts spatial resolution and image clarity. As the pixel size increases, both the generated 3D images and the given 2D image become progressively blurred. The tiny pores and unreacted clinker particles in the image become difficult to distinguish from the hydration products, leading to an increasing image-measured proportion of hydration products and a decreasing proportion of pores and unreacted clinkers. Thus, although the proposed method can synthesize realistic 3D images, high-resolution images with smaller pixel sizes are more conducive to accurately analyzing the proportions of various phases in cement.

\subsection{Comparison Experiment}
In this section, our method is evaluated through comparison with the current leading technique in 3D microstructure reconstruction, SliceGAN \cite{slicegan}. In the experiments, we compare and analyze two aspects, computation time and reconstruction quality. As shown in Figure \ref{duibiimg}, the two methods synthesize 3D microstructures based on the same 2D BSE image (266 × 266 pixels$ ^ 2$). The size of the 3D image generated by both methods is the same, 256 × 256 × 256 pixels$ ^ 3$. In terms of synthesis performance, the 3D microstructures generated by our method more closely resemble the real 2D image than those generated by SliceGAN, which indicates that our method has the advantage in reconstructing complex 3D microstructures.

\begin{table}[h]
\renewcommand{\arraystretch}{1.0}
\centering
\caption{Time comparison, h refers to hours and s refers to seconds.}
\label{timeduibi}
\begin{tabular}{ccc}
\hline
Methods  & Training time & Reconstruction time \\ \hline
Ours     & 15.8h         & 4s ($256 \times 256 \times 256$)                 \\
SliceGAN & 15.5h         & 1s ($256 \times 256 \times 256$)                 \\ \hline
\end{tabular}
\end{table}

Table \ref{timeduibi} presents a comparison between our method and SliceGAN in terms of training and reconstruction time. To ensure a fair comparison, both methods are tested on the same computational device. As shown in Table \ref{timeduibi}, our method performs slightly worse than SliceGAN in terms of computational time. Although our approach increases computational cost due to specific designs (e.g., the multi-scale learning strategy and network architecture), it ensures enhanced reconstruction quality.

\section{Conclusion}
\label{conclusion}
This paper proposes a GAN-based method for generating 3D microstructural images of hardened cement pastes. In the method, an improved image generation framework (CEM3DMG) is designed to generate 3D microstructural images based on the given 2D cross-sectional image.

\begin{itemize}
    \item The proposed CEM3DMG requires only a 2D BSE image, after which any number of 3D microstructural images are generated, providing comprehensive characterization and realistic microstructures for cement hydration. Compared to other 3D imaging techniques (such as FIB-SEM and synchrotron radiation-based Micro-CT), the economic cost is effectively controlled.

    \item The proposed CEM3DMG generates high-quality and realistic 3D microstructural images. In experiments, generated 3D microstructural images exhibit visual features similar to the given 2D image. Furthermore, the generated 3D images and the real 2D images show similarity in aspects such as gray level distribution, various phases proportions, and pore size distribution. Experimental results prove the effectiveness of the proposed method.

    \item The proposed CEM3DMG can generate large-size 3D microstructural images similar to the 2D image. In experiments, a subset of image with various sizes is verified, including 3D images of $200 \times 200 \times 200$ pixels$ ^ 3$, $400 \times 400 \times 400$ pixels$ ^ 3$, $800 \times 800 \times 800$ pixels$ ^ 3$. Moreover, a slice of $1000 \times 3000$ pixels$ ^ 2$ from large-size 3D images is exhibited.
\end{itemize}

The proposed 3D microstructural image generation method is a promising technology. However, there is still room for improvement. As a deep learning-based approach, one limitation is its dependency on the 2D image (scan area) size. Deep learning-based models typically rely on the volume and richness of information in the data. Larger scan area provides more detailed microstructure information, leading to improved reconstruction outcomes. Conversely, smaller scan areas might compromise the quality of the generated 3D microstructures. One potential solution to overcome this limitation is to acquire multiple 2D images from various regions of the sample and combine them for a more comprehensive reconstruction. Furthermore, as the method relies on deep learning, the training process is inherently time-consuming, although the lengthy training process can effectively improve the reconstruction quality. Future efforts will focus on designing more efficient network architectures and developing faster training methods to alleviate these challenges.

\section*{Data availability statement}
The source code and data for CEM3DMG are available in the GitHub repository at: https://github.com/NBICLAB/CEM3DMG.

\bibliography{mybibfile}

\end{document}